%Paper: hep-ph/9207217
%From: Boris Blok <BLOK@sbitp.itp.ucsb.edu>
%Date: Tue, 7 Jul 1992 00:30 PST

%%	JNL.TEX					Doug Eardley
%%
\message
{JNL.TEX version 0.95 as of 5/13/90.  Using CM fonts.}
%%
%%	This is a set of TeX 82 macros designed to produce scientific
%%	papers with a minimum of fuss and using as much of plain.tex as
%%	possible.  The user need only know what is in the TeXbook, and
%%	the macros under ``user definitions'' below.  Also, the user
%%	definitions are intended to be as simple as possible, so that
%%	the user may change them as desired.  I have tried to avoid all
%%	cleverness, although it may have snuck in here and there.
%%
%%	A considerable degree of compatibility with AmSTeX is maintained,
%%	although not guaranteed.  The intention is that AmSTeX input file
%%	should run with only a few changes near the beginning;  see
%%	discussion below under "AmSTeX compatability".
%%
%%	For documentation, see the file JNLHLP.TEX.  Optional features are
%%	contained in the files PPT.TEX (for two-up preprints), REFORDER.TEX
%%	(automatic numbering of references), EQNORDER.TEX (automatic numbering
%%	of equations), and TABLEOFC.TEC (automatic generation of table of
%%	contents).

%%
%%	Redefine \input to prevent files being loaded more than once
%%
\catcode`@=11
\expandafter\ifx\csname inp@t\endcsname\relax\let\inp@t=\input
\def\input#1 {\expandafter\ifx\csname #1IsLoaded\endcsname\relax
\inp@t#1%
\expandafter\def\csname #1IsLoaded\endcsname{(#1 was previously loaded)}
\else\message{\csname #1IsLoaded\endcsname}\fi}\fi
\catcode`@=12

%%
%%  Font definitions for Computer Modern (CM) Fonts
%%
\font\twelverm=cmr12			\font\twelvei=cmmi12
\font\twelvesy=cmsy10 scaled 1200	\font\twelveex=cmex10 scaled 1200
\font\twelvebf=cmbx12			\font\twelvesl=cmsl12
\font\twelvett=cmtt12			\font\twelveit=cmti12
\font\twelvesc=cmcsc10 scaled 1200	\font\twelvesf=cmss12
                     
%%\font\twelvemib=ambi10 scaled 1200
%%\font\tenmib=ambi10
%%\font\eightmib=ambi10 scaled 800
%%\font\sixmib=ambi10 scaled 667
\font\twelvemib=cmmib10 scaled 1200
\font\tenmib=cmmib10
\font\eightmib=cmmib10 scaled 800

\skewchar\twelvei='177			\skewchar\twelvesy='60
\skewchar\twelvemib='177
%  Define \...point macros to change fonts and spacings consistently

\newfam\mibfam

\def\twelvepoint{\normalbaselineskip=12.4pt plus 0.1pt minus 0.1pt
  \abovedisplayskip 12.4pt plus 3pt minus 9pt
  \belowdisplayskip 12.4pt plus 3pt minus 9pt
  \abovedisplayshortskip 0pt plus 3pt
  \belowdisplayshortskip 7.2pt plus 3pt minus 4pt
  \smallskipamount=3.6pt plus1.2pt minus1.2pt
  \medskipamount=7.2pt plus2.4pt minus2.4pt
  \bigskipamount=14.4pt plus4.8pt minus4.8pt
  \def\rm{\fam0\twelverm}          \def\it{\fam\itfam\twelveit}%
  \def\sl{\fam\slfam\twelvesl}     \def\bf{\fam\bffam\twelvebf}%
  \def\mit{\fam 1}                 \def\cal{\fam 2}%
  \def\sc{\twelvesc}		   \def\tt{\twelvett}%
  \def\sf{\twelvesf}               \def\mib{\fam\mibfam\twelvemib}%
  \textfont0=\twelverm   \scriptfont0=\tenrm   \scriptscriptfont0=\sevenrm
  \textfont1=\twelvei    \scriptfont1=\teni    \scriptscriptfont1=\seveni
  \textfont2=\twelvesy   \scriptfont2=\tensy   \scriptscriptfont2=\sevensy
  \textfont3=\twelveex   \scriptfont3=\twelveex\scriptscriptfont3=\twelveex
  \textfont\itfam=\twelveit
  \textfont\slfam=\twelvesl
  \textfont\bffam=\twelvebf \scriptfont\bffam=\tenbf
                            \scriptscriptfont\bffam=\sevenbf
  \textfont\mibfam=\twelvemib \scriptfont\mibfam=\tenmib
                              \scriptscriptfont\mibfam=\eightmib
  \normalbaselines\rm}

%	tenpoint

%%
%%	Change of the internal codes for lowercase Greek letters
%%	in order to use "${\mib\alpha}$"... for boldface "alpha"... .
%%      (by Ulrich Kettler 5/8/86)
%%
\mathchardef\alpha="710B
\mathchardef\beta="710C
\mathchardef\gamma="710D
\mathchardef\delta="710E
\mathchardef\epsilon="710F
\mathchardef\zeta="7110
\mathchardef\eta="7111
\mathchardef\theta="7112
\mathchardef\iota="7113
\mathchardef\kappa="7114
\mathchardef\lambda="7115
\mathchardef\mu="7116
\mathchardef\nu="7117
\mathchardef\xi="7118
\mathchardef\pi="7119
\mathchardef\rho="711A
\mathchardef\sigma="711B
\mathchardef\tau="711C
\mathchardef\phi="711E
\mathchardef\chi="711F
\mathchardef\psi="7120
\mathchardef\omega="7121
\mathchardef\varepsilon="7122
\mathchardef\vartheta="7123
\mathchardef\varpi="7124
\mathchardef\varrho="7125
\mathchardef\varsigma="7126
\mathchardef\varphi="7127

%%
%%	Various internal macros
%%

\def\beginlinemode{\endmode
  \begingroup\parskip=0pt \obeylines\def\\{\par}\def\endmode{\par\endgroup}}
\def\beginparmode{\endmode
  \begingroup \def\endmode{\par\endgroup}}
\let\endmode=\par
{\obeylines\gdef\
{}}
\def\singlespace{\baselineskip=\normalbaselineskip}

\def\oneandahalfspace{\baselineskip=\normalbaselineskip
  \multiply\baselineskip by 3 \divide\baselineskip by 2}
\def\doublespace{\baselineskip=\normalbaselineskip \multiply\baselineskip by 2}

\newcount\firstpageno
\firstpageno=2
\footline={\ifnum\pageno<\firstpageno{\hfil}\else{\hfil\twelverm\folio\hfil}%
\fi}
\def\toppageno{\global\footline={\hfil}\global\headline
  ={\ifnum\pageno<\firstpageno{\hfil}\else{\hfil\twelverm\folio\hfil}\fi}}
\let\rawfootnote=\footnote		% We must set the footnote style
\def\footnote#1#2{{\rm\singlespace\parindent=0pt\parskip=0pt
  \rawfootnote{#1}{#2\hfill\vrule height 0pt depth 6pt width 0pt}}}
\def\raggedcenter{\leftskip=4em plus 12em \rightskip=\leftskip
  \parindent=0pt \parfillskip=0pt \spaceskip=.3333em \xspaceskip=.5em
  \pretolerance=9999 \tolerance=9999
  \hyphenpenalty=9999 \exhyphenpenalty=9999 }
\def\dateline{\rightline{\ifcase\month\or
  January\or February\or March\or April\or May\or June\or
  July\or August\or September\or October\or November\or December\fi
  \space\number\year}}
\def\received{\vskip 3pt plus 0.2fill
 \centerline{\sl (Received\space\ifcase\month\or
  January\or February\or March\or April\or May\or June\or
  July\or August\or September\or October\or November\or December\fi
  \qquad, \number\year)}}

%%
%%	Page layout, margins, font and spacing (feel free to change)
%%

\hsize=6.5truein
\hoffset=0pt
%%\hoffset=1truein
\vsize=8.9truein
\voffset=0pt
%%\voffset=1truein
\parskip=\medskipamount
\def\\{\cr}
\twelvepoint		% selects twelvepoint fonts (cf. \tenpoint)
\doublespace		% selects double spacing for main part of paper (cf.
			%	\singlespace, \oneandahalfspace)
\overfullrule=0pt	% delete the nasty little black boxes for overfull box

%%
%%	The user definitions for major parts of a paper (feel free to change)
%%

  %  "Draft", Timestamp

	% Preprint number at upper right of title page

\def\title			%  Title on title page
  {\null\vskip 3pt plus 0.2fill
   \beginlinemode \doublespace \raggedcenter \bf}

\def\author			%  Author(s) name(s)  on title page
  {\vskip 3pt plus 0.2fill \beginlinemode
   \singlespace \raggedcenter\sc}

\def\affil			% Affiliations (can intermix with \author)
  {\vskip 3pt plus 0.1fill \beginlinemode
   \oneandahalfspace \raggedcenter \sl}

\def\abstract			% Begin abstract
  {\vskip 3pt plus 0.3fill \beginparmode
   \oneandahalfspace ABSTRACT: }

\def\endtitlepage		% End title page, begin body of paper
  {\endpage			% 	This subsumes \body
   \body}

\def\body			% Begin text body;  can be used to end
  {\beginparmode}		% \title, \author, \affil, \abstract,
				% \reference, or \figurecaption modes

\def\head#1{			% Head;  NOTE enclose the text in {}
  \goodbreak\vskip 0.5truein	%  e.g., \head{I. Introduction}
  {\immediate\write16{#1}
   \raggedcenter \uppercase{#1}\par}
   \nobreak\vskip 0.25truein\nobreak}

\def\itemitemitem{\par\indent\indent \hangindent3\parindent \textindent}
\def\itemitemitemitem{\par\indent\indent\indent \hangindent4\parindent
\textindent}
\def\beginitems{\par\medskip\bgroup
  \def\i##1 {\par\noindent\llap{##1\enspace}\ignorespaces}%
  \def\ii##1 {\item{##1}}%
  \def\iii##1 {\itemitem{##1}}%
  \def\iiii##1 {\itemitemitem{##1}}%
  \def\iiiii##1 {\itemitemitemitem{##1}}
  \leftskip=36pt\parskip=0pt}\def\enditems{\par\egroup}

\def\makefigure#1{\parindent=36pt\item{}Figure #1}

\def\figure#1 (#2) #3\par{\goodbreak\midinsert
\vskip#2
\bgroup\makefigure{#1} #3\par\egroup\endinsert}

\def\beneathrel#1\under#2{\mathrel{\mathop{#2}\limits_{#1}}}

\def\refto#1{$^{#1}$}		% For references in text as superscript

\def\references			% Begin references -- basic format is Phys Rev
  {\head{References}		% I.e., volume, page, year (space after commas).
   \beginparmode
   \frenchspacing \parindent=0pt \leftskip=1truecm
   \parskip=8pt plus 3pt \everypar{\hangindent=\parindent}}

\gdef\refis#1{\item{#1.\ }}			% Ref list numbers.

\gdef\journal#1, #2, #3, 1#4#5#6{		% Journal reference.  Comma sets
    {\sl #1~}{\bf #2}, #3 (1#4#5#6)}		% off: name, vol, page, year

\def\endreferences{\body}

\def\figurecaptions		% Begin figure captions
  {\endpage
   \beginparmode
   \head{Figure Captions}
}

\def\endpage			%  Eject a page
  {\vfill\eject}

\def\endpaper			%  Ways to say goodbye
  {\endmode\vfill\supereject}

%%
%%	AmSTeX compatability definitions
%%
%%	To run a TeX file originally intended for AmSTeX, only small changes
%%	should be necessary (I hope).  Use the line \input jnl at the start.
%%	Remove the lines \input amstex, \documentstyle{itpjnl} at the
%%	beginning;  also remove all the page layout stuff (\parindent=1cm,
%%	\hsize=5.28125in etc.)  The page layout is now done automatically.
%%	Also OMIT the qualifier \magnification=1200 when you IMPRINT the
%%	.dvi file.  (\TagsOnRight is harmless, you can take it out or leave
%%	it in.)  I believe most AmSTeX will work with no change.  One problem
%%	is \footnote, which is a little different in that it now needs to
%%	have an explicit asterisk *  (or whatever) included, like this:
%%		\footnote*{Text winds up at bottom of page.}
%%	This is discussed on p. 116 of the TeXbook.  IGNORE the AmSTeX
%%	documentation (if you can call it that);  refer to the TeXbook.
%%
%%	Note that many commands in AmSTeX have their equivalents in the
%%	TeXbook, perhaps with different names and slightly differing
%%	usage. E.g., the old \align in AmSTeX is replaced by \eqalign
%%	(p. 190) and \aligntag is replaced by \eqalignno (p. 192).
%%	\align and \aligntag still work, but I recommend that you use
%%	\eqalign and \eqalignno in documents run under jnl.
%%
%%	See me if you have any problems  -- Doug.
%%

\def\heading				% Heading
  {\vskip 0.5truein plus 0.1truein	% e.g., \heading I. NOTES \endheading
   \beginparmode \def\\{\par} \parskip=0pt \singlespace \raggedcenter}

\def\subheading				% Subheading
  {\vskip 0.25truein plus 0.1truein	% e.g., \subheading{A. The Problem}
   \beginlinemode \singlespace \parskip=0pt \def\\{\par}\raggedcenter}

\def\tag#1$${\eqno(#1)$$}

\def\align#1$${\eqalign{#1}$$}

\def\aligntag#1$${\gdef\tag##1\\{&(##1)\cr}\eqalignno{#1\\}$$
  \gdef\tag##1$${\eqno(##1)$$}}

\def\endaligntag{}

\def\overset #1\to#2{{\mathop{#2}\limits^{#1}}}
\def\underset#1\to#2{{\let\next=#1\mathpalette\undersetpalette#2}}
\def\undersetpalette#1#2{\vtop{\baselineskip0pt
\ialign{$\mathsurround=0pt #1\hfil##\hfil$\crcr#2\crcr\next\crcr}}}

%%
%%	Various little user definitions
%%

\def\ref#1{Ref.~#1}			% 	for inline references
\def\Ref#1{Ref.~#1}			% 	ditto
\def\[#1]{[\cite{#1}]}
\def\cite#1{{#1}}
			% For figure numbers
		% For citation of equation numbers
	%	ditto
			%	ditto
			%	ditto
		%	ditto
\def\(#1){(\call{#1})}
\def\call#1{{#1}}
\def\taghead#1{}
\def\frac#1#2{{#1 \over #2}}

\def\12{{1\over2}}

\def\sla{\raise.15ex\hbox{$/$}\kern-.57em}
\def\leaderfill{\leaders\hbox to 1em{\hss.\hss}\hfill}
\def\twiddle{\lower.9ex\rlap{$\kern-.1em\scriptstyle\sim$}}
\def\bigtwiddle{\lower1.ex\rlap{$\sim$}}
\def\gtwid{\mathrel{\raise.3ex\hbox{$>$\kern-.75em\lower1ex\hbox{$\sim$}}}}
\def\ltwid{\mathrel{\raise.3ex\hbox{$<$\kern-.75em\lower1ex\hbox{$\sim$}}}}
\def\square{\kern1pt\vbox{\hrule height 1.2pt\hbox{\vrule width 1.2pt\hskip 3pt
   \vbox{\vskip 6pt}\hskip 3pt\vrule width 0.6pt}\hrule height 0.6pt}\kern1pt}
\def\tdot#1{\mathord{\mathop{#1}\limits^{\kern2pt\ldots}}}
\def\happyface{%
$\bigcirc\rlap{\lower0.3ex\hbox{$\kern-0.85em\scriptscriptstyle\smile$}%
\raise0.4ex\hbox{$\kern-0.6em\scriptstyle\cdot\cdot$}}$}
\def\sadface{%
$\bigcirc\rlap{\lower0.25ex\hbox{$\kern-0.85em\scriptscriptstyle\frown$}%
\raise0.43ex\hbox{$\kern-0.6em\scriptstyle\cdot\cdot$}}$}

\def\pmb#1{\setbox0=\hbox{#1}%
  \kern-.025em\copy0\kern-\wd0
  \kern  .05em\copy0\kern-\wd0
  \kern-.025em\raise.0433em\box0 }

\catcode`@=11
\newcount\tagnumber\tagnumber=0

\immediate\newwrite\eqnfile
\newif\if@qnfile\@qnfilefalse
\def\write@qn#1{}
\def\writenew@qn#1{}
\def\w@rnwrite#1{\write@qn{#1}\message{#1}}
\def\@rrwrite#1{\write@qn{#1}\errmessage{#1}}

\def\taghead#1{\gdef\t@ghead{#1}\global\tagnumber=0}
\def\t@ghead{}

\expandafter\def\csname @qnnum-3\endcsname
  {{\t@ghead\advance\tagnumber by -3\relax\number\tagnumber}}
\expandafter\def\csname @qnnum-2\endcsname
  {{\t@ghead\advance\tagnumber by -2\relax\number\tagnumber}}
\expandafter\def\csname @qnnum-1\endcsname
  {{\t@ghead\advance\tagnumber by -1\relax\number\tagnumber}}
\expandafter\def\csname @qnnum0\endcsname
  {\t@ghead\number\tagnumber}
\expandafter\def\csname @qnnum+1\endcsname
  {{\t@ghead\advance\tagnumber by 1\relax\number\tagnumber}}
\expandafter\def\csname @qnnum+2\endcsname
  {{\t@ghead\advance\tagnumber by 2\relax\number\tagnumber}}
\expandafter\def\csname @qnnum+3\endcsname
  {{\t@ghead\advance\tagnumber by 3\relax\number\tagnumber}}

\def\equationfile{%
  \@qnfiletrue\immediate\openout\eqnfile=\jobname.eqn%
  \def\write@qn##1{\if@qnfile\immediate\write\eqnfile{##1}\fi}
  \def\writenew@qn##1{\if@qnfile\immediate\write\eqnfile
    {\noexpand\tag{##1} = (\t@ghead\number\tagnumber)}\fi}
}

\def\callall#1{\xdef#1##1{#1{\noexpand\call{##1}}}}
\def\call#1{\each@rg\callr@nge{#1}}

\def\each@rg#1#2{{\let\thecsname=#1\expandafter\first@rg#2,\end,}}
\def\first@rg#1,{\thecsname{#1}\apply@rg}
\def\apply@rg#1,{\ifx\end#1\let\next=\relax%
\else,\thecsname{#1}\let\next=\apply@rg\fi\next}

\def\callr@nge#1{\calldor@nge#1-\end-}
\def\callr@ngeat#1\end-{#1}
\def\calldor@nge#1-#2-{\ifx\end#2\@qneatspace#1 %
  \else\calll@@p{#1}{#2}\callr@ngeat\fi}
\def\calll@@p#1#2{\ifnum#1>#2{\@rrwrite{Equation range #1-#2\space is bad.}
\errhelp{If you call a series of equations by the notation M-N, then M and
N must be integers, and N must be greater than or equal to M.}}\else%
 {\count0=#1\count1=#2\advance\count1
by1\relax\expandafter\@qncall\the\count0,%
  \loop\advance\count0 by1\relax%
    \ifnum\count0<\count1,\expandafter\@qncall\the\count0,%
  \repeat}\fi}

\def\@qneatspace#1#2 {\@qncall#1#2,}
\def\@qncall#1,{\ifunc@lled{#1}{\def\next{#1}\ifx\next\empty\else
  \w@rnwrite{Equation number \noexpand\(>>#1<<) has not been defined yet.}
  >>#1<<\fi}\else\csname @qnnum#1\endcsname\fi}

\let\eqnono=\eqno
\def\eqno(#1){\tag#1}
\def\tag#1$${\eqnono(\displayt@g#1 )$$}

\def\aligntag#1\endaligntag
  $${\gdef\tag##1\\{&(##1 )\cr}\eqalignno{#1\\}$$
  \gdef\tag##1$${\eqnono(\displayt@g##1 )$$}}

\def\eqalignno#1{\displ@y \tabskip\centering
  \halign to\displaywidth{\hfil$\displaystyle{##}$\tabskip\z@skip
    &$\displaystyle{{}##}$\hfil\tabskip\centering
    &\llap{$\displayt@gpar##$}\tabskip\z@skip\crcr
    #1\crcr}}

\def\displayt@gpar(#1){(\displayt@g#1 )}

\def\displayt@g#1 {\rm\ifunc@lled{#1}\global\advance\tagnumber by1
        {\def\next{#1}\ifx\next\empty\else\expandafter
        \xdef\csname @qnnum#1\endcsname{\t@ghead\number\tagnumber}\fi}%
  \writenew@qn{#1}\t@ghead\number\tagnumber\else
        {\edef\next{\t@ghead\number\tagnumber}%
        \expandafter\ifx\csname @qnnum#1\endcsname\next\else
        \w@rnwrite{Equation \noexpand\tag{#1} is a duplicate number.}\fi}%
  \csname @qnnum#1\endcsname\fi}

\def\ifunc@lled#1{\expandafter\ifx\csname @qnnum#1\endcsname\relax}

\let\@qnend=\end\gdef\end{\if@qnfile
\immediate\write16{Equation numbers written on []\jobname.EQN.}\fi\@qnend}

\catcode`@=12

\catcode`@=11
\newcount\r@fcount \r@fcount=0
\newcount\r@fcurr
\immediate\newwrite\reffile
\newif\ifr@ffile\r@ffilefalse
\def\w@rnwrite#1{\ifr@ffile\immediate\write\reffile{#1}\fi\message{#1}}

\def\writer@f#1>>{}
\def\referencefile{%			  Stuff to write .REF file
  \r@ffiletrue\immediate\openout\reffile=\jobname.ref%
  \def\writer@f##1>>{\ifr@ffile\immediate\write\reffile%
    {\noexpand\refis{##1} = \csname r@fnum##1\endcsname = %
     \expandafter\expandafter\expandafter\strip@t\expandafter%
     \meaning\csname r@ftext\csname r@fnum##1\endcsname\endcsname}\fi}%
  \def\strip@t##1>>{}}

\def\citeall#1{\xdef#1##1{#1{\noexpand\cite{##1}}}}
\def\cite#1{\each@rg\citer@nge{#1}}	% Variable No. of args, separated by ","

\def\each@rg#1#2{{\let\thecsname=#1\expandafter\first@rg#2,\end,}}
\def\first@rg#1,{\thecsname{#1}\apply@rg}	% each@ag is a general purpose
\def\apply@rg#1,{\ifx\end#1\let\next=\relax%	  variable no. of arg. macro.
\else,\thecsname{#1}\let\next=\apply@rg\fi\next}% args separated by commas

\def\citer@nge#1{\citedor@nge#1-\end-}	% Check for M-N range (M and N numbers)
\def\citer@ngeat#1\end-{#1}
\def\citedor@nge#1-#2-{\ifx\end#2\r@featspace#1 % Single argument
  \else\citel@@p{#1}{#2}\citer@ngeat\fi}	% M-N range of arguments
\def\citel@@p#1#2{\ifnum#1>#2{\errmessage{Reference range #1-#2\space is bad.}%
    \errhelp{If you cite a series of references by the notation M-N, then M and
    N must be integers, and N must be greater than or equal to M.}}\else%
 {\count0=#1\count1=#2\advance\count1
by1\relax\expandafter\r@fcite\the\count0,%
  \loop\advance\count0 by1\relax%	  Loop from M to N
    \ifnum\count0<\count1,\expandafter\r@fcite\the\count0,%
  \repeat}\fi}

\def\r@featspace#1#2 {\r@fcite#1#2,}	% Eat spaces at beginning or end of arg
\def\r@fcite#1,{\ifuncit@d{#1}%		  Cite individual reference
    \newr@f{#1}%
    \expandafter\gdef\csname r@ftext\number\r@fcount\endcsname%
                     {\message{Reference #1 to be supplied.}%
                      \writer@f#1>>#1 to be supplied.\par}%
 \fi%
 \csname r@fnum#1\endcsname}
\def\ifuncit@d#1{\expandafter\ifx\csname r@fnum#1\endcsname\relax}%
\def\newr@f#1{\global\advance\r@fcount by1%
    \expandafter\xdef\csname r@fnum#1\endcsname{\number\r@fcount}}

\let\r@fis=\refis			% Save old \refis, redefine
\def\refis#1#2#3\par{\ifuncit@d{#1}%      Use two params #2 #3 to strip blank
   \newr@f{#1}%
   \w@rnwrite{Reference #1=\number\r@fcount\space is not cited up to now.}\fi%
  \expandafter\gdef\csname r@ftext\csname r@fnum#1\endcsname\endcsname%
  {\writer@f#1>>#2#3\par}}

\def\ignoreuncited{%   redefine \refis if ignoring uncited references
   \def\refis##1##2##3\par{\ifuncit@d{##1}%
     \else\expandafter\gdef\csname r@ftext\csname
r@fnum##1\endcsname\endcsname%
     {\writer@f##1>>##2##3\par}\fi}}

\def\r@ferr{\endreferences\errmessage{I was expecting to see
\noexpand\endreferences before now;  I have inserted it here.}}
\let\r@ferences=\references
\def\references{\r@ferences\def\endmode{\r@ferr\par\endgroup}}

\let\endr@ferences=\endreferences
\def\endreferences{\r@fcurr=0%		  Save old \endreferences, redefine
  {\loop\ifnum\r@fcurr<\r@fcount%	  Loop over refnum and produce text
    \advance\r@fcurr by 1\relax\expandafter\r@fis\expandafter{\number\r@fcurr}%
    \csname r@ftext\number\r@fcurr\endcsname%
  \repeat}\gdef\r@ferr{}\endr@ferences}

% Save old \endpaper, redefine it to write parting message.

\let\r@fend=\endpaper\gdef\endpaper{\ifr@ffile
\immediate\write16{Cross References written on []\jobname.REF.}\fi\r@fend}

\catcode`@=12

\citeall\refto		% These macros will generate citations
\citeall\ref		%
\citeall\Ref		%

\rightline{NSF-ITP-92-100}
\rightline{TPI-MINN-92-32/T}
\rightline{June  1992}
\vskip.8in
\centerline{\bf The Isgur-Wise function in the small velocity limit}
\bigskip
\centerline{B. Blok
%\footnote*{Research supported by
%DOE grant DE-FG02-90ER40542}
}
\bigskip
\centerline{\sl Institute for Theoretical Physics}
\centerline{\sl University of California at Santa Barbara}
\centerline{\sl Santa Barbara, CA 93106 }
\centerline{\sl and}
\centerline{M. Shifman
%$^*$
%\footnote*{$^*$Research supported by
%NSF grant DMS-8610730}
}
\bigskip
\centerline{\sl  Theoretical Physics Institute}
\centerline{\sl University of Minnesota}
\centerline{\sl Minnesota, MN 55455}
%\endpage
\bigskip
\abstract{ We discuss the Isgur-Wise function $\xi (y)$ in the small velocity
(SV) limit within the QCD sum rule method. The behavior of $\xi (y)$ in
the SV limit is sensitive to the particular form of the duality relations used
to decontaminate the sum rule predictions from the continuum contribution.
Peculiarities of the duality relations in the problem at hand are revealed.
It is shown that the proper requirements of duality and angular isotropy
for S wave states
 lead to an unambiguous form of the sum rules for the Isgur-Wise
function. We illustrate the constraints due to these requirements using a
toy model of the harmonic oscillator. The slope parameter and the shape of
$\xi (y)$ are determined.

}
\endpage
\head{1. Introduction.}
\par Strong interactions in the quark systems become simpler if one
 of the quarks considered is infinitely heavy. Then its motion is
 effectively classic and can be considered as given in the sense that the
 surrounding gluons and light quarks (we will refer to this component as to
the light cloud, which is equivalent to more commonly used
 notion of "brown muck") do not affect it at all. The fact that the simplest
hadronic system one can invent is built from an infinitely heavy quark plus
 the light cloud has been noted long ago . \refto{1} The revival of interest to
such systems we are witnessing now is due to explicit introduction of new
 symmetries \refto{2} taking place if there are several heavy quarks, and $m_Q
\gg \Lambda_{\rm QCD}$ where $m_Q$ is the heavy quark mass. The initial
 impetus in this direction has been given by the work  \refto{3} where the
$B\rightarrow D$ transition has been analysed in the so called small velocity
(SV) limit, see below. The culminating point in the recent achievements
in the "heavy" quark-"light" quark systems is the formulation of the Heavy
Quark Effective Theory \refto{4} (for a review see \ref{5}), which ideally
suits
for describing regularities of this sector of the hadron world.
\par In the semileptonic transitions of heavy mesons of different flavors the
heavy quark symmetry is reflected in the universal Isgur-Wise (IW) function.
Assume we are interested in the current induced transitions (it can be either
vector $V_\mu$ or axial $A_{\mu}$ current) between two heavy mesons,
$M_i$ and $M_f$. The quark content of $M_{i,f}$ is $Q_{i,f}\bar q$ where
$Q_i(Q_f)$ is the initial (final) heavy quark and $\bar q$ is a light
antiquark.
For instance, we can take b as $Q_i$ and c as $Q_f$. Moreover, assume for
definiteness  that $M_i$ is the ground state pseudoscalar (generic B). $M_f$
can be either the ground state pseudoscalar or the ground state vector
(generic $D$ or $D^*$). In general, the set of transition matrix elements of
$V_\mu$ and $A_\mu$ involves a large number of apriori unrelated formfactors.
In the limit $m_Q\rightarrow \infty$ all these matrix elements are described by
a single function which can be introduced in the following way:
$$<P_{ Q_f\bar q}(v)\vert V^\mu\vert P_{ Q_i\bar q}(v_i)>
=\sqrt{M_fM_i}\xi(y)
(v_f^\mu+v_i^\mu).\eqno (1)$$
Here $P_{ Q\bar q}$ stands for the pseudoscalar meson with the quark content
$ Q\bar q$, $V^{\mu}=\bar Q_f\gamma_\mu Q_i$, $M_f (M_i)$ denotes the mass of
the
 final (initial) meson, $v^\mu$ is the four-velocity,
$$v^\mu_{i,f}=P^\mu_{i,f}/M_{i,f},\eqno (2)$$
and, finally, the variable $y\equiv v_fv_i$. In the limit $m_Q\rightarrow
\infty$ we stick to the quark and meson masses coincide. The peculiarity of the
IW formfactor $\xi$ is that it depends only on the velocity transfer
( masses are irrelevant) \refto{2}. If in the rest frame of the initial meson
the final meson is also at  rest then $\xi$ is trivially known \refto{3}:
$$\xi(y=1)=1.\eqno (3)$$
The normalization property of the IW function in the SV limit is essentially
 not dynamical, it merely expresses the symmetry of the strong interaction.
 At the same time the shape of the function $\xi (y)$ encodes dynamical
 properties of Quantum  Chromodynamics. One can show that $\xi (y)$ is related
to the Wilson line  operator for specific contours and is, thus, one of
 the most fundamental objects in QCD. Dynamical QCD-based calculations of the
IW function are, clearly, of general interest.
\par Several recent works are devoted to this issue. In \ref{6,7,8,8a} the
IW function is treated within the QCD sum rule approach \refto{9}, with
 results that seem to be quite promising.
(The first applications of the QCD sum rules to the form factor problems
are worked out in \ref{701,702,23,25}).
 Still the question is far from being
closed since some aspects obviously require further study. One of the
most evident examples is the small velocity (SV) limit \refto{3} of the IW
 function. The slope of the function $\xi$ at $y\rightarrow 1$ reflects an
 interesting characteristics of the formfactor (see below). In \ref{7} it comes
out infinite, an unphysical feature of the corresponding analysis. A remedy
suggested in \ref{8} eliminates the spurious infinity, but leaves the
prediction
for slope parameter
 rather uncertain, to say nothing about an apparent theoretical
ambiguity contained in a prescription
used in  \ref{8} which calls for clarification.
\par The present paper is specifically devoted to the analyses of the IW
 function in the SV limit. First, we discuss different
specific aspects of the QCD sum
 rule calculations, elucidating new elements which go beyond the standard
procedure \refto{9}.
The main focus is on the issue of how duality must be applied to estimate
continuum contribution in the formfactor situation. In the SV limit at least
one of the standard prescriptions is shown to be contradictory and the
procedure
of the continuum decontamination should be carried out in a specific way.
 Second, we comment on the additional information stemming
from the Bjorken sum rule \refto{10} which relates the slope of $\xi$ in the SV
limit to nondiagonal formfactors of the type $Q_i\bar q\vert_{0^-}\rightarrow
Q_f\bar q\vert_{1^+}$.
\par Our main results can be summarized as follows. The duality for
the three-point
functions must be understood not in a local, but in a generalized sense,
as duality between integrated spectral densities. This is explained in detail
in section 2 on the example of a toy model of the harmonic oscillator.
The sum rules for the Isgur-Wise function and its slope
parameter $\rho^2$ are
described in section 3. We argue that the symmetry requirements and
the proper duality
lead to an unambiguous form of the sum rules. This, in turn, leads to
the elimination of uncertainties noticed in \ref{8,8a}. In section 4
we analyse the sum rules and get predictions for
the Isgur-Wise function at small
$y$ and for $\rho^2$.
 We also show the self-consistency of our approach using
the Bjorken sum rule.

\head{2. Calibrating the sum rule approach in the harmonic oscillator.}

\par In this section we shall test various elements of the method \refto{9}
in peculiar kinematics inherent to the problem at hand. To this end a toy
model is considered where there are no gluons and the
interaction between q and Q is pure potential. For simplicity we will choose
the
harmonic oscillator potential,
$$V(\vec r)={m\omega^2\vec r^2\over 2}\eqno (2.1)$$
where $\vec r$ is the distance between Q and q, m is a mass of the "light
quark".
\par It is convenient to work in the Breit reference frame ( Fig.1). Before the
photon emission the heavy quark force center moves with the velocity $\vec v$,
after the emission its velocity is $-\vec v$, so that the momentum transfer
q=$(0,\vec q)$, where $\vec q=2M\vec v$, $M\rightarrow \infty$ is
a mass of the heavy quark. Since our toy
model is nonrelativistic (with respect to the light quark q) we have to
 assume that $\omega \ll m$ and, moreover, $\vert \vec v\vert \ll \sqrt{
\omega /m}$.
The latter constraint is imposed because we are interested in the SV limit.
In other words the formfactor is expanded in powers of $\vec v$, and only the
terms $O(\vec v^0)$ and $O(\vec v^2)$
 are kept. The first term trivially reduces to unity
$-$an analog of eq. \(3)$-$while the coefficient in front of $\vec v^2$ is the
central object of our analysis.
\par Needless to say that in the potential model the diagonal formfactor can be
found explicitly and in the straightforward way, as well as nondiagonal
 transitions involved in the Bjorken sum rule. The nonrelativistic analogue of
the IW function has the form (in the Breit system)
$$\xi(\vec v^2)_{\rm n.r.}
=\int d^3x \psi^*_0(\vec r)e^{-2im\vec v\vec r}\psi_0(\vec r).
\eqno (2.2)$$
Here $\psi_0$ is the ground state wave function and and the subscript "n.r."
stands for "nonrelativistic".
Eq. \(2.2) implies that for the oscillator potential \(2.1)
$$\xi_{\rm n.r.}=1-{m\over \omega }\vec v^2+...\quad .\eqno (2.3)$$
In order to
make
contact with the standard parametrization accepted in the relativistic case
(see e.g. \refto{26}),
$$\eqalign{&\xi (y)=1-\rho^2 (y-1)+O((y-1)^2),\cr
&y=v'_\mu v^\mu ,\cr} \eqno (2.4)$$
it is convenient to present eq. \(2.3) as
$$\rho_{n.r.}^2={m \over 2\omega}.\eqno (2.5)$$
Here the fact that
 $$y-1=2\vec v^2+O(\vec v)^4\eqno (2.6)$$
is explicitly taken into account. Notice that for genuinely non-relativistic
systems parameter $\rho^2$ is large:
 ${m\over \omega}\gg 1$.
\par Along with the diagonal formfactor \(2.2) one can consider nondiagonal
transition formfactors from S to P-wave states, nonrelativistic analogues of
the
universal functions introduced in \ref{26}. In this case
$$2iv_j\tau_{\rm n.r}=\int d^3 x \psi^*_0(x)e^{-2im\vec v\vec r}
\psi_j(x)\eqno (2.7)$$
where $\psi_j(x)\quad (j=1,2,3)$ is the three-component wave function of the
P-wave state:
$$\psi_j =\sqrt{3\over 4\pi}{r_j\over \vert r\vert }R_1(r)\eqno (2.6a)$$
 The explicit forms of $\psi_0$ and $\psi_j$ can be found
in the standard textbooks( see e.g. \ref{12}). At small $\vec v^2$
the transition amplitude is
proportional to $\vec v$. In the general case there are transitions from
the ground state $\psi_0$ to all excited P-wave states. For the harmonic
oscillator, however, the only nonvanishing $\tau$ is that due to the transition
to the lowest P-wave state.
\par Now, the Bjorken sum rule \refto{10} takes the form:
$$\rho^2_{n.r.}=r^2_{\rm n.r.}(\vec v^2=0).\eqno (2.8)$$
The latter equation expresses the completeness of the full set of the
 wave functions. From eq. \(2.7) it trivially stems that
$$\tau (\vec v^2=0)=-\sqrt{m\over \omega}{1\over \sqrt{2}}\eqno (2.9)$$
Comparing eqs. \(2.5), \(2.8) and \(2.9)
 we see that the Bjorken sum rule is satisfied.
Of course, there is no doubt that it should be satisfied. We do this simple
exercise here  only because we want to set the framework and test the sum rule
approach, to be applied in more complex environment of quarks and gluons.
\par Let us pretend now that the wave functions are unknown, and the only
quantities we are able to calculate reliably  are the correlation functions
at short (euclidean) times. Specifically, we consider
$$\eqalign{S(\vec v;\tau_1,\tau_2)&=\int d^3rK(0,\tau_1+\tau_2\vert \vec r,
\tau_1)e^{-2im\vec v \vec r}
K(\vec r,\tau_1\vert 0,0)\cr
&\equiv \int d^3r\Sigma_l e^{-E_l\tau_2}
\psi_l(0)\psi_l^*(\vec r)e^{-im\vec v \vec r}
\Sigma_n
\psi^*_n(0)\psi_n(\vec r)e^{-im\vec v \vec r}e^{-E_n\tau_1}.\cr}\eqno (2.10)$$
This is the nonrelativistic analog of three point function. Here
$K(\vec r,\tau\vert \vec 0 , 0)$ is the amplitude  (time-dependent
Green function ) of the propagation from the point
 $(0,0)$ to the point $\vec r, \tau$
in the Euclidean time. If $\tau_{1,2}\rightarrow \infty$ only the ground state
contributes and $S(\vec v,\tau_1,\tau_2)\rightarrow e^{-E_0T}\vert
 \psi_0 (0)\vert^2\xi_{\rm n.r.}(\vec v^2),\,\,\,
 T=\tau_1+\tau_2$. For the harmonic
oscillator $K$ is known exactly for all $\tau$ (see e.g. \ref{13}),
$$K(\vec x,\tau\vert 0,0)=\big({m\omega\over 2\pi \sinh{\omega\tau}}
\big)^{3/2}e^{-{m\omega\over
 2\sinh{(\omega\tau)}}\vec x^2\cosh{(\omega\tau)}}.\eqno (2.11)$$
However we shall base our estimates on the short time expansion of K in order
to imitate QCD calculations, and then compare the results obtained in this
way with the exact ones.
\par Integration over $\vec r$ in eq. \(2.10) is readily carried out, and we
arrive at the following $\vec v^2$ expansion
$$S(\vec v;\tau_1 , \tau_2)=S_0+S_1+... \eqno (2.12)$$
where $S_i$ is the term of the $i-th$ order in $\vec v^2$,
$$S_0=({m\omega\over 2\pi})^{3/2}
{1\over (\sinh{\omega T})^{3/2}},\quad T=\tau_1+\tau_2 ;\eqno
(2.13)$$
$$S_1=({m\omega\over 2\pi})^{3/2}{1\over (\sinh{\omega T})^{3/2}}
{\sinh{(\omega\tau_1)}\sinh{(\omega\tau_2)}\over
 \sinh{\omega T}}{m\over \omega}(-2\vec v^2).\eqno (2.14)
$$
Notice that the zero-order term is just the same as for two-point functions
(see e.g. \ref{14}), a straightforward consequence of the normalization
theorem \refto{3}. It is well known that the sum rule approach gives beautiful
results for the positions of the S-wave levels in Quantum mechanics \refto{15},
a refined expression of duality which works in a very transparent way in this
 case. We will concentrate on the term $O(\vec v^2)$
determining the slope of the formfactor.
\par Combining eqs. \(2.14) and \(2.13) in the limit $\tau_{1,2}\rightarrow
 \infty$ one would recover eq. \(2.5) ( e.g. by considering the ratio
$S_1/S_0$.) We would like to examine how far one can go with the
short $\tau$ expansion.
\par Keeping only the first terms of the expansion already produces
 a reasonable estimate even disregarding the contribution due to higher states
( the latter is at the center of the discussion below). Indeed, in the
 symmetric point
$\tau_1=\tau_2=T/2$
$$S_1/S_0=(-2\vec v^2){m\over \omega }{x\over 4}
(1-{1\over 12}x^2+...),\quad x \equiv \omega T .
\eqno (2.15)$$
It is easy to see that the stability plateau is around $x\sim 2$;
 the corresponding estimate for $\rho^2$ is $\rho^2\sim {m\over 2\omega}(2/3)$
(cf. eq. \(2.5)). At this point the power correction constitutes 1/3 of the
 leading term. Moreover, with 3 power terms included $\rho^2\sim
 (m/(2\omega))0.8
$. Agreement with the exact result \(2.5) is evident, although not
 very impressive, of course, which is natural, since in the above analysis
 we completely ignored contamination due to higher states.
\par Let us discuss now how the higher states affect the analysis. $S_0$
is contributed by the S-levels only, situated at $E_n={3\over 2}\omega
 +2n\omega$, only
diagonal transitions take place,
$$S_0=\sum^{\infty}_{n=0}\vert \psi_n(0)\vert^2e^{-E_nT},\eqno (2.15a)$$
where
$$\eqalign{\vert \psi_n(0)\vert^2 &
=({m\omega\over \pi})^{3/2}{(2n+1)!!\over 2^nn!}\cr
&\rightarrow\sqrt{{E_n\over \omega}}({m\omega\over \pi})^{3/2}\sqrt{2\over \pi}
\quad {\rm at} \quad E_n\gg\omega .\cr
}
\eqno (2.16)$$
Eq. \(2.16) perfectly matches  the "bare" spectral density, i.e. that
 determining $S_0$ in the limit when the interaction is switched off,
$\omega \rightarrow 0$,
$$\sigma_0^{\rm bare}={1\over \sqrt{2\pi}}({m\over \pi})^{3/2}\sqrt{E}\,\,\,\,,
S_0^{\rm bare}=\int^{\infty}_0\sigma^{\rm bare}_0e^{-ET}dE.\eqno (2.17a)$$
Indeed, the smeared sum of delta-functions, $\sigma=\sum_n\sqrt
{{2\over \pi}}\sqrt{{E_n\over \omega}}({m\omega\over \pi})^{3/2}\delta
(E-E_n)$,
coincides with eq. \(2.17a) . The standard step-like model \refto{9} with
$\sigma_0$ from eq. \(2.17a) and the threshold energy $E_0={5\over 2}\omega$
gives excellent approximation for higher states \refto{15}.
\par Let us proceed to
 less studied case of $S_1$ where, unlike $S_0$,
we do encounter peculiarities
 of the form-factor situation. From the large $\tau$
expansion of eq. \(2.14) we learn that $S_1$ contains both the diagonal and
off-diagonal transitions,
$$\eqalign{S_{1diag}&=\sum^{\infty}_{k=0}({-2\over 3}m^2\vec v^2)(r^2_{k,k})
\vert \psi_k(0
)\vert^2e^{-E_kT}\cr
&=(-2\vec v^2)({m\over \omega})({m\omega\over 2\pi})^{3/2} ({1\over
 \sinh{\omega T}}
)^{5/2}{1\over 2}\cosh{\omega T}\cr},\eqno (2.17)$$
and
$$\eqalign{S_{1off-diag.}&=\sum^{\infty}_{k=0}(-{2\over 3}
 m^2\vec v^2)(r^2)_{k,k+1}
\psi_k(0)\psi^*_{k+1}(0)e^{-E_kT}(e^{-2\omega\tau_2}+e^{-2\omega\tau_1})\cr
&=(-2\vec v^2)({m\over \omega})({m\omega\over 2\pi})^{3/2}{1\over
 (\sinh{\omega T})^{5/2}}
(-1/4)e^{\omega T}(e^{-2\omega\tau_2}+e^{-2\omega\tau_1}).\cr}\eqno (2.18)$$
Here
$$r^2_{l,k}=\int d^3r\psi^*_l(\vec r)\vec r^2\psi_k(\vec r),\eqno (2.19)$$
and we have used the fact specific for the oscillator that the off-diagonal
 transitions are only from $l$ to $l\pm 1$ S wave state. Thus, the ground state
is connected only to the first excited S-wave state. $\psi_k(\vec r)$ stands
 for the wave function of the k-th S-wave state. The set of points where
 the physical spectral density is concentrated is presented in Fig. 2.
\par Now, one can easily convince oneself that the residue of the k-th state
contributing to $S_{1diag}$ $(E_k={(3/2)\omega +2\omega k})
$ is equal to
$$(-{2\over 3}m^2\vec v^2)(r^2)_{k,k}\vert \psi_k(0)\vert^2=-2\vec v^2({m\over
2
\omega})({m\omega\over \pi})^{3/2}\{{(4k+3)\over 3}{1\over 2^{2k}}
{1\over (k!)^2}(2k+1)!\}.\eqno (2.20)$$
Furthermore, for the off-diagonal transitions, eq. \(2.18), the residues are
$$(-{2\over 3} m^2\vec v^2)(r^2)_{k,k+1}\psi_k(0)\psi_{k+1}^*(0)
=-2\vec v^2({m\over 2\omega})({m\omega\over \pi})^{3/2}(-1)
\{{2k+3\over 3}{1\over (k!)^2}{(2k+1)!\over 2^{2k}}\}.\eqno (2.21)$$
Notice the important minus sign appearing in passing from eq. \(2.20) to
eq. \(2.21) .
\par At this point it becomes clear that the standard picture of
 local duality we got used to in the case of the sum rules for the two-point
functions \refto{9} (see e.g. \ref{16} )  fails in the formfactor situation,
 at least in the kinematics at hand (expansion of the formfactor near
the SV limit).
Indeed, the standard picture would imply that a few individual levels smeared
over a natural duality domain produce (locally) the same spectral density
inside this domain as the one obtained from free  motion. We will see shortly
that a conventional averaging of eqs. \(2.20) and \(2.21) leads to something
 quite
different from the "bare" spectral density, even in sign.
\par First we need to calculate the function $S(\vec v; \tau_1,\tau_2)$ defined
in eq. \(2.10) for the case of free motion ($\omega =0$). This is a
 straightforward exercise since
in this case
$$S_{bare}=\int\int {d^3\vec pd^3\vec p'\over (2\pi)^6}
 d^3\vec r e^{i\vec p\vec r
-i\vec p'\vec r}\exp{\{-{\vec p^2\over 2m}\tau_1-{\vec p^{'2}\over
2m}\tau_2\}};
\eqno (2.22)$$
$\vec r$ is the distance between the particle and the force center; $
{\vec p^2\over 2m}$ and
${\vec p^{'2}\over 2m}$ are energies.
\par We begin with a free particle with the momentum $\vec p'+m\vec v$ in the
Breit frame ($\vec p'$ is in the rest frame of the force center Q, see Fig.1),
and end up with the
 particle with the momentum $\vec p-m\vec v$ in the Breit frame
($\vec p$ in the rest frame of the force center Q, after it emits a quanta
"$\gamma$"  which carries away $\vec q=2M\vec v$, Fig.1). Since there is
 no interaction, nothing happens to the particle in the process of scattering
 of the force center,
$$\vec p'+m\vec v=\vec p-m\vec v.\eqno (2.23)$$
Eq. \(2.23), obvious as it is, also formally stems from eq. \(2.22) which can
be
readily transformed to
$$S_{\rm bare}={1\over 8\pi^2}{m\over \vert \vec v\vert}\int_{\Sigma}
dEdE'e^{-E\tau_1-E'\tau_2}\eqno (2.24)$$
where the integration in the right-hand side runs over the domain
$\Sigma$ depicted in Fig.3.

In the limit $v\ll 1$ we are interested in the bare spectral density
$$\sigma_{\rm bare}(E)={1\over 8\pi^2}{m\over \vert\vec v\vert}\eqno (2.25)$$
is very large and is concentrated in the narrow strip near the diagonal $E=E'$
(Fig.3). If, instead of E and $E'$, one introduces the variables
$$A=E'-E,\quad B={1\over 2}(E'+E)\eqno (2.26)$$
the boundary of $\Sigma $ is given by the equation
$$B={1\over 8m\vec v^2}A^2+{1\over 2} m\vec v^2.\eqno (2.27)$$
Moreover, B varies in the interval
$[mv^2/2,\infty)$; for given B the maximal and minimal values of
 A are $\pm (B-{mv^2\over 2})^{1/2}(8mv^2)^{1/2}$. Thus, at $v\rightarrow 0$
the
width of the shaded strip in Fig. 3$-$ it is proportional to v $-$ vanishes.
Naively, one would say that the bare spectral density is dual to diagonal
transitions only (We hasten to add that the latter statement is not correct).
The fact that $\sigma_{\rm bare} >0$ will be important in what follows.
\par One more observation important for understanding how the duality can be
 implemented in the case at hand. The process with the interaction switched off
which must be dual to that with the bound state propagation looks as follows.
The propagation of the force center before and after scattering is like the
cannon ball, (Fig.1) and we need not consider it. The propagation of the
light "particle" is free. In the laboratory (Breit) frame we fix its energy,
otherwise allowing it to propagate freely in any direction it wants from the
point of emission to the point of absorbtion. Fixing the energy of the light
 particle
in the Breit frame means that we fix
$$B={1\over 2}(E+E')={\cal E}+{mv^2\over 2}\eqno (2.28)$$
where the energy in the Breit frame is denoted as ${\cal E}$.
 On the other hand,
allowing for all possible angles for the vector $\vec p_0=\vec p'+m\vec v=
\vec p-m\vec v$
 implies that $A=E-E'$ must be allowed to take all kinematically possible
values. Thus it is clear that the domains of duality can be
 only those depicted on Fig. 4.
Any other orientation of the duality domains by no means can be acceptable, and
as a matter of fact, leads to unphysical result for the slope parameter as will
be discussed in more detail below.
\par Let us now return to the problem of duality and discuss it in more detail.
Consider first the spectral density for the free particle. This density is
 concentrated in a small domain of the width $\sim \vert\vec v\vert\sqrt{m{\cal
E}}$
around the diagonal and is positive.
 The Taylor coefficients of expansion of $S_{\rm bare}$
in $ v^2$ arise through expanding the boundary of the
duality domain in $\vec v^2$ and expanding the integration
variables.
The duality domain should be chosen as a certain interval in ${\cal E}$.
{}From eq. \(2.27)
 we see that for fixed ${\cal E}$ $A_{\rm max}$ and $A_{\rm min}$
have no $v^2$ expansion. The only $v^2$ dependence comes from eq. \(2.28)
when one expresses B in terms of ${\cal E}$
and $v^2$ and, then, expands
 in $v^2$. To illustrate this point it is instructive
to consider $S_{\rm bare}$ at $\tau_1=\tau_2=T/2$. Eq. \(2.24) implies then
that
$$S_{\rm bare}(\tau_1=\tau_2=T/2)={1\over 8\pi^2}{m\over v}\int^\infty_0
d{\cal E}\int^{A_{\rm max}}_{A_{\rm min}}dA\exp{-({\cal E}+mv^2/2)T}.\eqno
(901)
$$
The width of the $A$ integration is $4\sqrt{2m{\cal E}}v$. (As has been
already mentioned, this width scales as $\sim (m\omega)^{1/2}v\rightarrow 0$
at $v\rightarrow 0$). As a result,
$$S_{\rm bare}(\tau_1=\tau_2={T\over 2})={1\over \sqrt{2}\pi^2}m^{3/2}
\int^\infty_0d{\cal E}{\cal E}^{1/2}\exp{-({\cal E}+mv^2/2)T}\eqno (901)$$
which coincides, of course, with the small $T$ limit of eqs.
\(2.13) , \(2.14) , being expanded in $v^2$. The $v^2$ term in the expansion,
we are most interested in, in particular, takes the form
$$\eqalign{
S_1&=-mv^2T{1\over 2\sqrt{2}\pi^2}m^{3/2}\int^{\infty}_0d{\cal E}{\cal E}^{1/2}
\exp{-{\cal E}T}\cr
&=-mv^2{1\over 4\sqrt{2}\pi^2}m^{3/2}\int^{\infty}_0d{\cal E}{\cal E}^{-1/2}
\exp{-{\cal E}T}.\cr}\eqno (*)$$
The last line in this expression contains no explicit $T$ dependence
except the exponential.
\par
 On the other hand, consider the genuine
spectral density for the
 oscillator. The spectral density for the Taylor coefficients in the expansion
of S in powers of $ v^{2n}$ is spread over the length of $2n\omega$ from
the diagonal (Fig. 2). The reason is that it is proportional to matrix
elements of $r^{2n}$. For n=1 we see that the spectral density is spread
over the length $2\omega$ as depicted in Fig.2. Consequently, we see that
the spectral density for $S(\vec v^2;\tau_1,\tau_2)$ is spread over the whole
plane. The area in which it is concentrated has no correlation with the area
where the free particle density is concentrated. The boundary
of the area where the free particles spectral density is located
is
 determined by ${\cal E} $ and $\vec v^2$, the analogous boundary for
oscillator is infinite
(or determined by $\omega$ for Taylor coefficients). Clearly,  due to
this reason the oscillator spectral density can not be locally dual to the
 spectral density for free particles in the usual sense.
% Second important point is that it is easy
%to see that in the direction perpendicular to  diagonal spectral
%density of oscillator and Taylor coefficients quickly change signs
%(recall different signs in eqs. (2.20), (2.21)). The free particle density
%does not change the sign.
 We conclude that the local duality between the
spectral densities for the oscillator and for the free particles is absent
in the formfactor situation considered here.
\par Although the local duality is absent, there still remains a
weaker version of duality, namely a generalized duality.
The oscillator  spectral density
integrated in the A-direction
 will be matched by the spectral density for the
free particles integrated over the corresponding region, provided the weight
function in the ${\cal E}$ direction is exponential.
\par Let us
explain this assertion in more detail.
 Start with $S_0-$ the Taylor coefficient at zero order in the expansion
of S in powers of $\vec v^2$. For the oscillator this density is proportional
to $\sqrt{{\cal E}}$ and is concentrated on the diagonal. For free particles
 this is
a constant spread over a narrow
 area depicted in Fig.4. Let us now integrate the
free particle spectral density
 between the boundaries in the direction orthogonal to the diagonal
(here and below we shall call this direction the A-direction). We immediately
get the spectral density $\sim \sqrt{\cal E}$. We conclude that in this case
the oscillator density is dual to free particle density integrated
over the permitted interval in A direction. Note that the latter coincides
with the free particle density from the sum rules for the 2-point function
(see eq. \(2.17)). We thus find the origin of the duality for $S_0$ discussed
above (see eqs. \(2.15)-\(2.17),  where we discussed the duality for
$S_0$). Already for $S_0$ the duality is, rigorously speaking, a generalized
 one.
\par Consider now $S_1$.
We have already discussed above the bare spectral density. The integral
 representation
for $S_1$ has the integrand $\sim v^{-1}$ and a very narrow integration domain,
whose width in the A direction is $\sim v$. One can introduce an effective
spectral density in the $B$ (${\cal E}$) direction performing explicitly
integration over $A$. Then the singularity in $v$ is eliminated and the
correct analytic properties in $v^2$ are realised in a trivial way. One
 can read off the relevant expression for the effective spectral density from
eq. \(*) .
\par Now, proceed to the analysis of $S_1$ from the other side, with the
 interaction
switched on, $\omega \ne 0$. In this case our starting point is eqs.
\(2.17) , \(2.18) , \(2.20) , \(2.21) . A sketch of the spectral density
 is given
on Fig.2 where the k-th energy level is given by $E_k=(3/2+2k)\omega $.
Our task is to check that the generalized duality still takes place:
the appropriately defined sum over k matches eq. \(*) .
 In order to check duality we must go to the limit
$\omega/E_k\rightarrow 0$ and check that in this case the expansion becomes
that for the free particle. Here however we encounter a difficulty. The series
\(2.17), \(2.18) converge absolutely only when $\omega T$ is kept
finite. Otherwise
we encounter alternating series
for $S_{\rm diag}$ and $S_{\rm non.-diag.}$
 with the terms increasing like $k^{3/2}$.
If the contribution of the k-th diagonal transition is positive that of the
neighboring off-diagonal transitions is negative. Generally speaking, such
sum depends on the way of summation. The regularization of the sum is provided
by the exponential factor $\exp{-E_kT}$ in eqs.
\(2.17), \(2.18) . This factor prompts us how to carry out the summation.
\par Let us consider the symmetric point $\tau_1=\tau_2=T/2,\quad
T\rightarrow 0 \quad (k$ large). It is easy to see that the k-th state
contributing to $S_{1\rm diag}$ has the residue proportional to
$$k^{3/2}+\alpha k^{1/2}+\beta k^{-1/2}+...\eqno (910)$$
while that contributing to $S_{1\rm off-diag}$ has the residue
$$k^{3/2}+\tilde \alpha k^{1/2}+\tilde \beta k^{-1/2}+...\quad  .
\eqno (911)$$
In terms of energies
%The sum of these terms is clearly dependent on the way of summation.
%We want to find the way of summation such that the corresponding spectral
% density smoothly becomes that of free particle \(2.29) in the limit $\omega
%\rightarrow 0$. Consider first the diagonal part. Changing the variables
%from k to $E_k$ and going to big k it is easy to check

$$S_{1diag}(E_k)\sim \sum_k (2({E_k\over \omega})^{3/2}+\alpha ({
E_k\over \omega})^{1/2}+...)
{1\over \sqrt{\omega}}\exp{(-E_kT)}.\eqno (2.32)$$
%Here a is a numerical constant. We used the Wallis formulae:
%$${(2n-1)!!\over (2n)!!}\sim {1\over \sqrt{\pi}\sqrt{n}}(1-{1\over 8n}+....)
%\eqno (2.31)$$
For the nondiagonal case we get in the same way:
$$\eqalign{S_{1\rm non.-diag.}(E_k)\sim &-\sum_k (2({E_k\over \omega})^{3/2}+
\tilde \alpha ({ E_k\over \omega})^{1/2}+
\tilde \beta ({\omega\over E_k})^{1/2}+...)\exp{(-E_kT)}\cr
&{1\over
\sqrt{\omega}}(\exp{(-2\omega\tau_1)}+ \exp{(-2\omega\tau_2
)}).\cr}\eqno (2.33)$$
Here $\alpha , \beta ,
\tilde \alpha , \tilde \beta $ are numerical constants.
Notice that the weight factors in eqs. \(2.32) , \(2.33) are different.
 It is clear that the dangerous
terms in the expansion are those proportional to $(E_k/\omega)^{3/2}$,
$(E_k/\omega)^{1/2}$. Otherwise the series
 will smoothly go into \(*) in the limit $\omega\rightarrow 0$.
\par Let us discuss the dangerous terms in more detail. Eqs. \
\(2.17) , \(2.18), \(2.20), \(2.21) can be combined in the following way:
$$\eqalign{&
S_1=-2v^2({m\over 2\omega})({m\omega\over \pi})^{3/2}e^{-(3/2)(\omega T)}\cr
&\sum ^{\infty}_{k=0}{1\over 2^{2k}}{1\over (k!)^2}{(2k+1)!\over 3}
e^{(-2k\omega T)}\{(4k+3)-(4k+6)e^{-\omega T}\}.\cr} \eqno (j)$$
The parameter $T$ and characteristic values of k in the sum are not
independent: the choice of $T$ dictates what values of k saturate the sum.
The duality formally is to hold at large k (small $T,\quad T\sim k^{-1}+
Ck^{-2}+...$). Then, one can expand the expression in braces in $T$,
$$\eqalign{\{...\}&=(4k+3)-(4k+6)(1-\omega T+{\omega^2T^2\over 2}+...)\cr
&=-3+4k\omega T+...\quad .\cr}\eqno (n)$$
Notice that in $O(T^0)$ the terms $\sim k$ cancel. This means that under
the appropriate summation the terms $\sim (E_k/\omega)^{3/2}$ in eqs.
\(2.32) and \(2.33) annihilate each other.
\par The next step is to inspect what happens at the next-to-leading order.
Since the individual contributions grow like $k^{3/2}$ it is quite obvious
that the sum is saturated by terms with $k={3\over 4T\omega }$
(one can see this from maximizing $k^{3/2}\exp{(-2k\omega T)}$. As a result,
$-3$ in eq. \(n) is cancelled by $4k\omega T$. In terms of eqs. \(2.32)
and \(2.33) this fact implies cancellation of $O((E_k/\omega)^{1/2})$
parts.
\par As a result  we are left with terms $O((E_k/\omega)^{-1/2})$.
To see that the coefficient matches that in eq. \(*) one must keep the term
$T^2$ in eq. \(n) as well to use a relation between $T$ and $k$,
$T={3\over 4\omega k}+Ck^{-2}$, truncated at $O(k^{-2})$ level.
\par Notice  that our way of summation of diagonal and nondiagonal parts
precisely corresponds to the integration in the A direction.
We conclude that the spectral density  integrated in the A direction
 for the oscillator
is dual to the spectral density for free particles, also integrated in the A
direction. Notice that duality is valid not for local densities $\sigma(E,E')$,
but only for integrated densities in the A direction $\sigma ({\cal E})$.
It is also obvious, from the discussion above, that the integration in the A
direction is the only way to ensure duality. Otherwise we encounter
dangerous terms $\sim 1/\omega$ and/or can even get terms with the
wrong analytic properties ( not expandable in $\vec v^2$).\refto{7}
\par These considerations directly affect the possible choice of the
continuum model. It is clear that the considerations above lead to the choice
of the continuum model depicted on Fig.4 as the only possibility
{}.
 \par We discussed above an analogue of the double Borel transformed
  sum rules. During the investigation of the 3-point functions some authors
considered also the so called sum rules in the external field
\refto{21,22,23,24}, that correspond
to single Borel  transformation. A non-relativistic analogue of the single-
Borel transformed QCD sum rules is the function
$$S^{sin}= \sum_n(S)_{nn}\vert\psi_n(0)\vert^2\exp{-E_nT}-
\sum_{n,l}S_{nl}{\exp{(-E_n T )}-\exp{(-E_l T )}\over T (E_n-E_l)}
\psi_n(0)\psi_l^*(0).
\eqno (2.38)$$
Here $$S_{nl}=\int d^3\vec r\psi^*_n(\vec r)\exp{(-2im\vec v\vec r)}
\psi_l(\vec r).\eqno
 (2.39)$$
It is clear that this function differs from the function S given by
eq. \(2.10) (taken at $\tau_1=\tau_2$) by different weight for the
nondiagonal transitions. The single Borel sum rules
(we shall use this name instead of more lengthy single time
Borel transformed sum rules) are widely used, e.g. in
the study of the pion and proton formfactors \refto{21,22,23,24}.
 This is the reason why we
consider them here.
It is easy to see that in the limit $T\rightarrow\infty$
$$S^{sin}(\vec v^2 , T)\rightarrow (\vert\psi (0)\vert^2\xi_{\rm n.r.}
(\vec v^2)+
{{\rm const}\over \omega T})e^{-E_0 T }.\eqno (2.39a)$$
Here ${\rm const}=-\psi_0(0)\psi^*_1(0)S_{01}+...$
 measures the contribution of the
nondiagonal transitions.
Asymptotically at $T\rightarrow \infty$
$$\omega T S^{sin}(\vec v^2, T)=(a_0 x+b)e^{-E_0T}\eqno (2.40)$$
where $x=\omega T$ and $a_0=\vert\psi (0)\vert^2\xi_{\rm n.r.}(\vec v^2)$.
{}From eqs. \(2.17) , \(2.18)
 it is easy to find exactly the corresponding coefficient for
the terms $\sim \vec v^2$ in $S^{sin}$:
$$S^{sin}_1(T)=-2\vec v^2{m\over \omega}({m\omega\over 2\pi})^{3/2}
{1\over (\sinh{\omega T})^{5/2}}{1\over 2}(\cosh{\omega T}-
{\sinh{\omega T}\over \omega T}).\eqno (2.41)$$
The r.h.s. of eq. \(2.41) multiplied by $(\sinh{(\omega T)})^{3/2}
\omega T$ is well approximated
by a straight line \(2.40) for $x=\omega T $ for sufficiently large x.
Indeed, it has the form
$${S_1^{sin}\omega T\over S_0}=(-2v^2){m\over \omega}{1\over 2}
(x\coth{(x)}-1).\eqno (931)$$
Taking several first terms in the expansion around point $x=0$ we have:
$${S_1^{sin}\over S_0}=(-2v^2){m\over \omega}
(x^2/6-{1\over 90}x^4+...).\eqno (932)$$
The r.h.s. of eq. \(932) is well approximated by a straight line for $x\ge 2$.
We get $\rho^2\sim {m\over 2\omega}0.7$. Taking 3 power correction terms
we get $\rho^2\sim {m\over 2\omega}0.8$. This can be compared with
eq. \(2.15) and the discussion that followed. We see that
in the case of the harmonic oscillator the single Borel sum rule
works as good as double Borel sum rule.
\par We note however, that in the case of the electromagnetic formfactor
of pion the single Borel sum rules give overestimated value of
the formfactor and underestimated value of the
 slope parameter, although within the
error bars (see \ref{21,22} ).
\par Consider now the duality for the single Borel sum rule,
i.e. function
$S^{sin}$.
$S_{1\rm diag}$ is the same as in eq. \(2.17) . The off-diagonal part is
almost the same as in eq. \(2.18) , with $\tau_1=\tau_2=T/2$. The only
difference is that $e^{-2\omega\tau_1}+e^{-2\omega\tau_2}=2e^{-\omega T}$
in this equation is substituted by $(1-e^{-2\omega T})/(\omega T)$
in eq. \(2.41) . These two expressions have identical first two terms in
the $\omega T$ expansion. In accordance with the discussion above this means
  that the dangerous terms in
 $S^{sin}$ are precisely the same as the dangerous terms in the double
Borel sum rule considered above. The duality considerations are the same
for both sum rules. When we write this sum rule in terms of the
spectral density $\sigma ({\cal E}
)$ we do it after implicit integration in A
direction.
\par Let us now summarize lessons that follow from the study of duality for
three point functions for harmonic oscillator. Note that these lessons refer
not only to the case
at hand discussed
in the remainder of this article
, but also to other cases of the calculation of three-point
functions in QCD in similar kinematics.
First, considerations of duality and appropriate symmetries impose strict
constraints on the continuum model
 which must correspond to the region of Fig. 4 .
Second, duality is not local. The duality in three- point functions is
between integrated in the A direction spectral densities.

\head{3. Sum rules for the Isgur-Wise function.}

\par In this section we shall write the sum rules for
the IW function and extend the analysis of the preceding
section to the heavy quark case.
\par We begin from the double Borel sum rules for the
IW function in the symmetric point $\sigma_1=\sigma_2=2\sigma$,
where $\sigma$ denotes the Borel parameter. This sum rule
was already written in \Ref{6,7,8,8a}. We refer the reader to
these articles as well as to \Ref{1} for the general discussion
of Borel transformed nonrelativistic sum rules in heavy quark
systems and the derivation of the sum rule for the IW
function. We have rechecked that the coefficients in front of condensate
terms are correct.  The bare spectral density can be obtained from
calculations in \ref{30} which were carried out for finite heavy quark masses.
Using explicit expressions in that work it is not difficult to take the limit
$m_Q\rightarrow\infty$.
\par The sum rule for the IW function has the form:
$$\eqalign{&{3\over 8\pi^2(y+1)\sqrt{y^2-1}}
\int\int_{\Sigma}ds_1ds_2(s_1+s_2)\exp{-{(s_1+s_2)\over 2\sigma}}\cr
&-a/4+{b(2y+1)\over
 2^63\sigma^2}={F^2\xi(y)\exp{-E_R/\sigma}\over 4}.\cr}\eqno (3.6)$$
Here $a,b$ are the vacuum condensates averages:
$$\eqalign{&a=<\bar \psi(0)\psi (0)>_0=-(0.24)^3\quad {\rm GeV}^3;\cr
&b=<\bar\psi (0) gG_{\mu\nu}\sigma_{\mu\nu}\psi (0)>_0=m_0^2a\,\,;\quad
m_0^2\sim 0.8\quad {\rm GeV}^2,\cr}\eqno (3.6ab)$$
$y$ is the recoil parameter, $\sigma$ is the Borel parameter.
\par Our parameters are  connected with the analogous parameters
used in \ref{8} in the following way. The Borel parameter $T$ used in \ref{8}
is $T=2\sigma$.
The spectral density variables
used in our eq. \(3.6) are connected with the dimensionless spectral variables
used in eq. (3.14) in \ref{8} as
$s_1=z\sigma , \quad s_2=z'\sigma$.
 The continuum threshold in that paper $\omega_c$ is
numerically
twice as big as the continuum threshold $\omega_c$ used in our paper,
since the author uses the spectral density variables $\omega ,\omega'$
equal  to $2s_1$, $2s_2$ respectively.
\par  Recall that
in the nonrelativistic case we make the Borel transform over the
light quark virtual energy and $\sigma$ has the dimension of mass,
not of mass square as in the relativistic case. This makes
the sum rules very similar to the sum rules for the harmonic oscillator
considered in the previous section. The area of the integration
of the spectral density is denoted by $\Sigma$ and is depicted in Fig. 5.
Finally, $E_R$ is the mass difference between the mass of the heavy
quark and that of the heavy meson (both tend to infinity in our limit).
The constant F is the heavy meson leptonic decay constant:
$$<0\vert \bar q\gamma_\mu\gamma_5 h_Q\vert P>
={iF\over \sqrt{2}}v_\mu\eqno (3.6a)$$
The field $h_Q$ are heavy quarks fields in Heavy Quark Effective
Theory \refto{4}. The constant $F$ is related to more familiar $f_P$ by
$F=f_P\sqrt{m_Q} $ modulo logs of the type discussed in \ref{31} .
\par  Although the sum rule is the same, our analysis will
be quite different from the analysis in \ref{7,8,8a}, however. In particular
we shall argue that the uncertainty in $\rho^2$ is quite small,
not gigantic as in the latter papers.  Indeed,
the main unsolved problem in the analysis of the sum rule \(3.6) is
the choice of continuum.
We saw in the previous section that the considerations due to duality
and relevant symmetry severely constrain the possible choice of
continuum model in the sum rules. The analysis in the previous
section
can be exactly repeated for the nonrelativistic sum rules in QCD.
As a result the only possible choice of the integration area is that
depicted in Fig. 5.
For every fixed value of $(s_1+s_2)/2$ we have to integrate over all
possible values of $A=(s_1-s_2)$. Any other prescription for integration
of spectral density
will be incompatible with the general requirements of
 duality and angular isotropy (S wave states).
The variable $(s_1+s_2)/2$ plays the same role as ${\cal E}$ in the previous
section.
\par We have argued  in a very similar kinematics that
 duality in the three point functions must be understood  in
the generalized sense.
 Namely, the spectral density
for the interacting case integrated in the A direction matches that for the
free particle only after integration in the A direction.
  The same
kind of generalized duality must also hold for the QCD sum rules.
\par The generalized duality
 means that we have to write the sum rule for the spectral density already
integrated in the A direction (recall that this is the direction orthogonal
to the main diagonal).
This effective spectral density will be dependent only on the
 variable $(s_1+s_2)/2$.
 Only after this integration
 we can speak about duality and make a choice
of continuum model.
We now have the only one possible choice of the  continuum
model. The continuum
is approximated by the spectral density due to the bare loop, integrated
over the variable $A=(s_1-s_2)$. This is exactly the same continuum model
as the model that was used in the two-point sum rules.
Numerically, of course, it does not matter whether we assume local
duality and take the bare loop contribution  integrated over a two-dimensional
area depicted on Fig.5 as the continuum model or if we
approximate the A integrated continuum contribution by the bare
spectral density, integrated in the A direction,
 starting from some threshold. However our experience with
the harmonic oscillator shows, that only the second prescription is physically
correct.
\par We can now write a sum rule for the three point function based only on the
 generalized duality discussed above.
It has the form:
$${3\over (y+1)^2
\pi^2}\int^{\omega_c}_0s^2e^{-s/\sigma}ds-{a\over 4}+{b(2y+1)\over 192\sigma
^2}=
{F^2\xi (y)\exp{-E_R/\sigma}\over 4}.\eqno (4.1)$$
Here $s=(s_1+s_2)/2$.
This sum rule coincides with one of many possible forms of the sum rules
(due to different choices of the integration domain $\Sigma$ in eq. \(3.6))
considered in \ref{8}.
 Now we know that eq. \(4.1) is the only possible form of
the sum rule.
We also write the sum rule for the slope parameter $\rho^2$ directly:
$$(+{3\over 4\pi^2}\int^{\omega_c}_0s^2e^{-s/\sigma}
ds-{b\over 96\sigma^2}){\exp{E_R/\sigma
}\over
 F^2/4}
=\rho^2.\eqno (4.5)$$
\par Consider now the single Borel sum rule.
  This sum rule
was found to work as well as the double Borel sum rule
in the case of the harmonic oscillator.
 We test this sum rule
 in the heavy quark case.
\par Recall the basic idea behind this approach. We write the dispersion
relation for the polarization operator
$$\Pi (E_1,E_2,y)=\int^{\infty}_{0}{{\rm Im}
\Pi(s_1,s_2,y)\over (s_1+E_1)(s_2+E_2)}
ds\eqno (2.71)$$
where $$(v_1^\mu+v_2^\mu)\Pi(E_1,E_2,y) =\int d^4xd^4y\exp{i(k_1x+kz)}
<j_5(x)j_\mu(z)j^{+}_5(0)>.\eqno (2.81)$$
In eq. \(2.81) $k_1$ is the momentum of the initial heavy meson, $k$ is the
transferred momentum, $k_2=k_1+k$ is the momentum of the final heavy meson;
$k_\mu=(m_Q+E_1)v_{1\mu}$ and $k_{2\mu}=(m_Q+E_2)v_{2\mu}$. The current
 $j_5=\bar Q\gamma_5 q
 $ is the quark current that creates heavy
 mesons, and $j_\mu=\bar Q\gamma_\mu Q$ is a  vector current.
Q  and q are the heavy and light quark fields
respectively. In the double Borel transformed
sum rules we make the double Borel transformation of the sum rule \(2.71)
with respect to the
the variables $E_1$ and $E_2$
independently and saturate the r.h.s. of eq. \(2.71) by the
resonance contributions. In the single Borel transformed sum rules
 we put  $E=E_1=E_2$ from the very beginning
in eq. \(2.71) and then carry out the
  Borel transformation over the variable E. As it was
 explained in the previous section in the example of the
 harmonic oscillator  the
 difference between the double and single
 Borel transformed sum rules is that we take
the contribution of nondiagonal transitions with  different weights.
The spectral density for the bare loop now depends only on one variable $E$.
As it was explained above when we use such spectral density
we do it after an implicit
 integration in the A direction. Thus, the two sum rules are very
similar in spirit. The single Borel transformed sum rule has the form:
$$\Pi(x,y,\omega_c)={\xi(y) F^2\exp{(-E_R/\sigma)}\over 4}+
a(y)\sigma .\eqno (2.91)$$
Here $a(y)$ measures the contribution of nondiagonal transitions.
One can easily calculate $\Pi$ in this case taking into account the same
operators as we took into account writing the  double Borel sum rule.
The result is
$$\eqalign{&{3\over 4\pi^2}
{1\over y+1}(1+{1\over \sqrt{y^2-1}}\log{(y+\sqrt{y^2-1}}))
\int^{\omega_c}_0s^2e^{-s/\sigma}ds\cr
&-a/4+{b(4y+5)\over 576\sigma^2}
={F^2\over 4}(\xi (y)+a(y)\sigma)\exp{(-E_R/\sigma )}\cr}\eqno (3.1)$$
\head{4. Analysis of the sum rules.}
\par Let us now analyse our sum rules. We start from the sum rule \(4.1).
In order to minimize the
 effects related to the uncertainty in $F$, $E_R$ we deal
with the sum rule \(4.1) divided by the sum rule for $F$,$E_R$ (\ref{1,7,8,8a}
)
:
$$\xi (y)={I(\sigma ,y,\omega_c)\over I_M(\sigma ,\omega_c)}\equiv G(\sigma
,y,\omega_c).\eqno
(5.1)$$
Here
$$I(\sigma ,y,w_c)={3\over (y+1)^2\pi^2}
\int^{\omega_c}_0s^2e^{-s/\sigma}ds-{a\over 4}+{b(2y+1)\over 192\sigma ^2}
\eqno (5.1a)$$
and
$$I_M(\sigma ,\omega_c)={3\over 4\pi^2}\int^{\omega_c}_0s^2e^{-s/\sigma}
ds-{a\over 4}
+{b\over 64\sigma^2}.\eqno (5.2)$$
The function $I_M(\sigma,\omega_c)$ is the theoretical part of the sum rule for
the mass
and leptonic decay constant of heavy meson.
$$I_M(\sigma,\omega_c)={F^2\over 4}\exp{(-E_R/\sigma)}\eqno (5.3)$$
The continuum threshold is chosen to be the same as for the two-point sum
rules.
We did not find any significant dependence of the continuum threshold on
the recoil $y$. We depict the behaviour of $\xi (y)$ for different values
of the
 continuum threshold in Fig.6. We find only a small uncertainty $\sim 10\%$
 due to the choice of the continuum threshold.
Let us also remind once again that we assume that the generalized duality
picture is valid. Hence our continuum model looks like the continuum
model for the sum rules for the two point function and has a single
threshold $\omega _c$. We vary this threshold to obtain the best stability of
the sum rules.
\par Let us now proceed to the sum rules for $\rho^2$, eq. \(4.5).
The dependence of the r.h.s. of this sum rule on $\sigma$ is depicted in Fig.7.
We immediately get:
$$\rho^2=0.7\pm 0.1.\eqno (5.8)$$
\par Consider now the single Borel transformed sum rules. We rewrite this
sum rule in the form:
$$a(y)\sigma+\xi (y)={I(\sigma
,y,\omega_c)\over I_M(\sigma ,\omega_c)}.\eqno (5.1ab)$$
Here the function $I$ is the l.h.s. of the sum rule \(3.1).
The function $a(y)$ is the $x-$independent function that measures
 the contribution of the nondiagonal transitions.
 The l.h.s.
of the sum rule \(5.1ab) is well approximated by a straight line for $\sigma\ge
0.8$ GeV. The corresponding behaviour of the Isgur-Wise function is depicted in
Fig.8.
Notice that the sum rule \(5.1a) gives slightly smaller ($\sim 10\%$ at
$y=2$) values of $\xi (y)$ than the sum rule \(4.5) . Nevertheless,
the difference between the predictions of these two sum rules is inside
error bars for each of them.
\par We can
 now differentiate the sum rule \(5.1a) with respect to $y$ at $y=1$.
Then we get the sum rule for $\rho^2$. This sum rule gives the value
$$\rho^2= 0.55\pm 0.1\eqno (5.2a)$$
Thus we get a slightly smaller value of $\rho^2$ using the sum rule \(5.1a).
The difference is, however, within the error bars.
\par It is tempting to speculate that the true value of $\rho^2$ lies
in the intersection of the predictions of the single and double
Borel sum rules.
 After all, these two sum rules worked equally well in the case of the harmonic
oscillator. This gives $\rho^2\sim 0.6-0.65$.
\par For more careful  estimate, we take the average of the predictions of
these two some rules.
We have $$\rho^2= 0.65\pm 0.15\eqno (941)$$
 as the average from the predictions of  these two sum rules.
\par Let us now turn to the check of self-consistency of our approach. The
nontrivial check is provided by  comparison of our results with the predictions
of
the Bjorken
sum rule \refto{10}.
\par The Bjorken sum rule \refto{10} has the form
$$\rho^2={1\over 4}+\sum_{m=1}^{m_{max}}\vert \tau_{1/2}^{m}(1)\vert^2
+2\sum_{m=1}^{p_{max}}\vert \tau_{3/2}^{m}(1)\vert^2+....\eqno (5.12)$$
Here $\tau_{1/2}(1)$, $\tau_{3/2}(1)$ are  nondiagonal transition
formfactors from the multiplet of pseudoscalar and vector mesons into
multiplets
with $s_l^{\pi_l}=1/2^+$ and $3/2^+$ respectively taken at zero recoil.
Here $s_l$ and $\pi_l$ are the spin and parity of the light quark in the meson.
These formfactors are defined in the following way:
$${<A\vert A_\mu\vert P>
\over m}\vert_{y=1}=-4\tau_{1/2}(v_1-v_2)_\mu\eqno (5.13)$$
$${<B\vert A_\mu\vert P>\over m}\vert_{y=1}
=-{1\over \sqrt{2}}i\tau_{3/2}\epsilon_{
\mu\alpha\beta\gamma}\epsilon^{*\alpha}(v_1+v_2)^\beta(v_1-v_2)^\gamma\eqno
 (5.14)$$
Here $A$  is the scalar heavy meson in $1/2^+$ multiplet, and $B$
 denotes the state with $J^P=1^+$ in $3/2^+$ multiplet of heavy mesons.
 We use the notations
 of \ref{25,26}. For general values of recoil $y$ the formfactors $\tau_{1/2}$,
$\tau_{3/2}$ are new universal functions of
the recoil analogous to the Isgur-Wise
function. All other formfactors for transitions between $1/2^-$ and
$1/2^+,3/2^+$ multiplets can be expressed through $\tau_{1/2}$ and
$\tau_{3/2}$.
\refto{26}
\par It is possible to analyse the formfactor $\tau_{1/2}(1)$
using   the QCD sum rules in which we do
 double Borel
transformation.
Taking into account only the contribution of the unit operator and the
quark condensate we obtain the following sum rule for $\tau_{1/2}$:
$$\eqalign{&
\lim_{y\rightarrow 1}{3\over 4\pi^2
\sqrt{y^2-1}(y-1)}\int_{\Sigma}(s_1-s_2)\exp{(-s_1/\sigma_1-s_2/\sigma
_2)}ds_1ds_2-\cr
&a/2=8\tau_{1/2}{\lambda_S \lambda_P\over 4}
\exp{(-E_S/\sigma_1-E_P/\sigma_2)}\cr}
\eqno (961)$$
A rough estimate using this sum rule gives
$$\tau_{1/2}(1)\sim 0.35\pm 0.2\eqno (5.18)$$
Note very high uncertainty in this estimate. First,
  we took into account only two operators
of dimensions zero (unit operator)  and 3 (light quark condensate).
Second, and more important, we did not take into account continuum
contribution.
In eq. \(961) $\lambda$
 is a semileptonic decay constant of the corresponding resonance:
$$<0\vert j\vert S(P)>=\lambda_{S(P)}\phi_{S(P)}\eqno (5.19)$$
where $j$ is the quark current that creates a meson ,
 $\bar Q q$ for scalar meson S
and $\bar Q\gamma_5q$ for pseudoscalar meson P,  $\phi$ is the
corresponding wave function.
\par It is interesting to note that the same ratio
 $\tau_{1/2}/\xi(1)$ can be obtained by combining nonrelativistic
quark model and predictions obtained using the
nonrelativistic sum rules for the two-point functions.\refto{1}
 Indeed, in the nonrelativistic quark  model\refto{26}
$$\tau_{1/2}(1)\sim \int d^3k\phi^*_{11}(k)k\phi_{00}(k)\eqno (5.20)$$
Here $\phi_{11}$ is the radial wave function of the scalar resonance, and
$\phi_{00}$-of pseudoscalar meson. For the rough estimate we have
$$\tau_{1/2}(1)={\tau_{1/2}(1)\over \xi(1)}
={\vert\phi_{11}(0)\vert\over \vert \phi_{00}(0)\vert}=\sqrt{{\lambda_P\over
\lambda_S}}\sim 0.3\eqno (5.21)$$
The latter equation follows from  \refto{1}
$$\lambda^2=12\vert\phi(0)\vert^2\eqno (5.22)$$
\par The determination of $\tau_{3/2}$ is much more cumbersome since the
 corresponding currents contain derivatives. However, following the
 nonrelativistic quark model we can accept for orientation
 that $\tau_{3/2}\sim \tau_{1/2}$. Note also that the contribution
of the radially excited states to Bjorken sum rule is strongly suppressed
(it is zero in nonrelativistic quark model).
We now take for $\tau_{1/2}$ the value (see eqs. \(5.18),\(5.21))
$\tau_{1/2}(1)\sim \tau_{3/2}(1)\sim 0.35$.
Then we get for $\rho^2$ an estimate:
$$\rho^2\sim{1\over 4}+3\tau_{1/2}(1)^2\sim 0.25 +3(0.35\pm 0.2)^2\eqno
(5.23)$$
We finally conclude that Bjorken sum rule implies
$$0.35\le \rho^2\le 1.15\eqno (5.24)$$
This result is in agreement with the sum rule prediction
\(5.8) .
\head{5. Conclusion}

\par We have used the QCD sum rule method
to find the shape of the Isgur-Wise function and its slope parameter
 $\rho^2$. We have
discussed the nature of the uncertainties pointed out in \ref{8,8a} and
shown that they can be eliminated if we  implement
the general requirements of
 duality
 in the appropriate way. We  used a toy model of a
harmonic oscillator to discuss in detail the consequences of the demand
of duality and relevant symmetries for the sum rules for formfactors.
\par In particular, we
have shown on the  example of harmonic oscillator
that there is no local duality in the sum rules for the three point
functions. Duality must be understood in a generalized sense, between
integrated quantities. The proper duality  severely
constrains the form of the continuum model.
\par For the case of heavy mesons the requirements of duality
 lead to the double Borel sum rules \(4.1), \(4.5). The corresponding behaviour
of the Isgur-Wise function as a function of recoil is depicted in Fig.6.
For the slope $\rho^2$ we get
$$\rho^2=+0.7\pm 0.1\eqno (6.1)$$
The single Borel transformed sum rules
lead to the shape of Isgur -Wise function depicted on Fig. 8. They lead to
the slightly smaller (within $10\%$) value of $\xi (y)$. The difference between
the predictions of these two sum rules is within the error bars however.
For the slope parameter we obtain
$$\rho^2= 0.55\pm 0.1\eqno (7.1)$$
For $\rho^2$ we can  argue that the values \(7.1) and \(6.1)
are upper and low bounds on the value of the slope parameter.
Taking the average of these two estimates , we have
$$\rho^2= 0.65\pm 0.15\eqno (7.2)$$
\par Note that our predictions for the slope parameter lead to the smaller
values than those obtained by analysing the phenomenological data:
$\rho^2\sim 1.4\pm 0.4$ in \ref{311} and $\rho^2\sim 1.2\pm 0.4$ in \ref{312}.
\par On the other hand, our results for
the slope parameter are very close to the results of \ref{25}. The model
developed in the latter paper gives the
value $\rho^2\sim 0.7$.
\par After this work was completed we have learnt about the results for
$\rho^2$ obtained from the so called optical sum rules \refto{313} extending
the Bjorken approach.\refto{10} The analysis in \ref{313}
leads to the inequality $\rho^2\le 0.75$. This is in a very good agreement
with our estimate for $\rho^2$ given by eqs. \(6.1) , \(7.1), \(7.2).
\par We are grateful for to A. Vainshtein and I. Kogan for useful discussions.

\endpage
\references

\refis{1} E. Shuryak, Nucl. Phys., B198 (1982) 83.

\refis{3} M. Voloshin and M. Shifman, Yad. Fiz.,
47(1988) 801 [Sov. J. of Nucl. Phys., 47 (1988) 511 ].

\refis{5} M. Wise, New symmetries of strong interactions, preprint CALT
68-1721,\hfill\break
 H. Georgi, Heavy quark effective field theory, preprint HUPT-91-A039,
1991;\hfill\break
 B. Grinstein, Light-Quark, Heavy-Quark systems, preprint SSCL-34,1992
(to be published in Ann. Rev. Nucl. Part. Sci.).

\refis{2} N. Isgur and M. Wise, Phys. Lett., B232 (1989) 113;
B237 (1990) 527.

\refis{701} B. L. Ioffe and A. V. Smilga, Phys. Lett., B114 (1982) 353.

\refis{702} A. V. Nesterenko and A. V. Radyushkin, Phys. Lett., B115 (1982)
410.

\refis{4} E. Eichten and B. Hill, Phys. Lett., B234 (1990) 511\hfill\break
H. Georgi, Phys. Lett., B240 (1990)447.

\refis{7} A. Radyushkin, Phys. Lett.,B271 (1991) 218.

\refis{6} M. Neubert, V. Rieckert, B. Stech and Q.P. Xu
, Heidelberg preprint HD-THEP -91-28 (1991).

\refis{8} M. Neubert, Phys. Rev., D45 (1991) 2451
{}.

\refis{10} J.D. Bjorken, preprint SLAC-PUB-5278 (1990), Invited talk at
Les Recontre de la valle d'Aosta, La Thuille, Italy, (1990).

\refis{12} S. Flugge, "Practical Quantum Mechanics", Springer-Verlag, 1965.

\refis{13} R. Feynman and A. Hibbs, "Quantum Mechanics and Path Integrals"
McGraw-Hill, 1965.

\refis{14} M. Shifman, Ann. Rev. Nucl. Part. Sci., 33 (1983) 199.

\refis{15} J. Bell and R. Bertlmann, Nucl. Phys., B177 (1981) 218;\hfill\break
A. I. Vainshtein et al, Yad. Fiz.,32 (1980) 1622 [Sov. J. of Nucl. Phys.,
32 (1980) 840].

\refis{16} B. V. Geshkenbein and M. S. Marinov, Yad. Fiz., 30 (1979)
1400 [Sov. J. of Nucl. Phys., 30 (1979) 726  ].

%\refis{4} M. Bauer, B. Stech and M. Wirbel, Z. Phys., C34 (1987) 103.

\refis{9} M. Shifman, A. Vainshtein and V. Zakharov, Nucl. Phys.,
B147
 (1979) 385, 448.

\refis{8a} P. Ball, Phys. Lett., B281 (1992) 133.

\refis{21} V.A. Nesterenko and A. I. Radyushkin, JETP Lett., 39 (1984) 707.

\refis{22} M. Beilin and A.I. Radyushkin, Int. J. of Mod. Phys., A3 (1988)
1183.

\refis{23}B. L. Ioffe and A. V. Smilga, JETP Lett., 37 (1983) 298.

\refis{24} I.I. Balitsky and A. Yung, Phys. Lett., 129B (1983) 388.

\refis{25} N. Isgur, D. Scora, B. Grinstein, M. Wise, Phys. Rev., D39 (1989)
 799;\hfill\break
B. Grinstein, M. B. Wise and N. Isgur, Phys. Rev. Lett., 56 (1986) 258.

\refis{26} N. Isgur and M. Wise, Phys. Rev., D43 (1991) 819.

\refis{31} M. Voloshin and M. Shifman, Yad. Fiz.,45 (1987) 463
[Sov. J. Nucl. Phys., 45 (1987) 292].

\refis{30} V. Eletsky and I. Kogan, Z. Phys., C28 (1985) 155.

\refis{313} M. Voloshin, preprint TPI-MINN-92/25-T (May 1992).

\refis{311} J.L. Rosner, Phys. Rev., D42 (1990) 3732.

\refis{312} M. Neubert, Phys. Lett.,B264 (1991) 455.

\endreferences

\endpage

\centerline{\bf Figure Captions}
\bigskip
{\bf Fig.1:} Heavy quark in Breit reference frame a) before and b) after the
               emission .

{\bf Fig.2:} The spectral density for $\rho^2_{\rm n.r.}$ for the oscillator.
The difference between $\bullet $ and 0 is the sign of the residue.

{\bf Fig.3:} Boundaries of the domain where the spectral density
 $\sigma_{\rm bare}\ne 0$.

{\bf Fig.4:} The only possible geometry of the duality domains.

{\bf Fig.5:} The duality domain for the sum rule for Isgur-Wise function.

{\bf Fig.6:} The Isgur-Wise function $\xi (y)$.

{\bf Fig.7} The dependence of the l.h.s. of the sum rule for $\rho^2$ on
the Borel parameter $x$.

{\bf Fig. 8} The Isgur-Wise function from the single Borel transformed sum
rule.

\endpage
\endpaper
\end\bye